\documentclass[letterpaper]{article} 
\usepackage{aaai2026}  
\usepackage{times}  
\usepackage{helvet}  
\usepackage{courier}  
\usepackage[hyphens]{url}  
\usepackage{graphicx} 
\usepackage{natbib}  
\usepackage{caption} 
\usepackage{algorithm}
\usepackage{algorithmic}
\usepackage{amsmath}
\usepackage{multirow}
\usepackage{colortbl}
\usepackage{booktabs} 
\usepackage{xcolor}
\usepackage{amssymb}
\usepackage{newfloat}
\usepackage{listings}
\DeclareCaptionStyle{ruled}{labelfont=normalfont,labelsep=colon,strut=off} 
\floatstyle{ruled}
\newfloat{listing}{tb}{lst}{}
\floatname{listing}{Listing}
\usepackage{url} 
\usepackage{cite}
\usepackage{amsthm,amsmath,amssymb}
\usepackage{mathrsfs}
\usepackage{array}
\usepackage{tikz}
\usepackage[skins]{tcolorbox} 
\tcbuselibrary{breakable} 
\usepackage{booktabs}  
\usepackage{multirow}
\usepackage{graphicx}
\usepackage{adjustbox}
\usepackage{colortbl}
\usepackage{bbding} 

\usepackage[capitalize]{cleveref}

\title{When Top-ranked Recommendations Fail: Modeling Multi-Granular Negative Feedback for Explainable and Robust Video Recommendation}
\author{
    Siran Chen\textsuperscript{\rm 1,\rm 2,\rm 3}\thanks{Equal contribution.},
    Boyu Chen\textsuperscript{\rm 1,\rm 2,\rm 3}\footnotemark[1],
    Chenyun Yu\textsuperscript{\rm 4}\footnotemark[2], 
    Yi Ouyang \textsuperscript{\rm 3},
    Lei Cheng \textsuperscript{\rm 3},
    Chengxiang Zhuo \textsuperscript{\rm 3},
    Zang Li \textsuperscript{\rm 3},
    Yali Wang\textsuperscript{\rm 1,\rm5}\thanks{Corresponding author.}
}
\affiliations{
    \textsuperscript{\rm 1} Shenzhen Key Laboratory of Computer Vision and Pattern Recognition, Shenzhen Institutes of Advanced Technology, Chinese Academy of Sciences, Shenzhen, China\\
    \textsuperscript{\rm 2} University of Chinese Academy of Science, Beijing, China \\
    \textsuperscript{\rm 3} Tencent, Shenzhen, China \\
    \textsuperscript{\rm 4} Shenzhen Campus of Sun Yat-sen University, Shenzhen, China\\
    \textsuperscript{\rm 5} Shanghai Artificial Intelligence Laboratory, Shanghai, China \\
    
    {chensiran17, chenboyu18}@mails.ucas.ac.cn,
    yuchy35@mail.sysu.edu.cn,
    yl.wang@siat.ac.cn
}

\usepackage{bibentry}

\begin{document}

\maketitle

\begin{abstract}

%
Existing video recommendation systems, relying mainly on ID-based embedding mapping and collaborative filtering, often fail to capture in-depth video content semantics. Moreover, most struggle to address biased user behaviors (e.g., accidental clicks, fast skips), leading to inaccurate interest modeling and frequent negative feedback in top recommendations with unclear causes.
To tackle this issue, we collect real-world user video-watching sequences, annotate the reasons for users' dislikes, and construct a benchmark dataset for personalized explanations. 
We then introduce the Agentic Explainable Negative Feedback (ENF) framework, which integrates three core components: 
(1) the Profile Agent, extracting behavioral cues from users' historical data to derive psychological and personality profiles; 
(2) the Video Agent, performing comprehensive multimodal video analysis; 
and (3) the Reason Agent, synthesizing information from the other two agents to predict user engagement and generate explanations.
Additionally, we propose the S-GRPO algorithm, enabling the model to progressively address complex tasks during reinforcement fine-tuning. 
Experimental results on the collected dataset show that our method significantly outperforms state-of-the-art baselines in negative feedback prediction and reason explanation.
Notably, it achieves an 8.6\% improvement over GPT-4o in reason classification.
Deployment on the business platform further validates its benefits: increasing average user watch time by 6.2\%, reducing the fast-skip rate by 9.4\% , and significantly enhancing user satisfaction.

\end{abstract}

\section{Introduction}

\begin{figure*}
    \centering
    \includegraphics[width=0.95\linewidth]{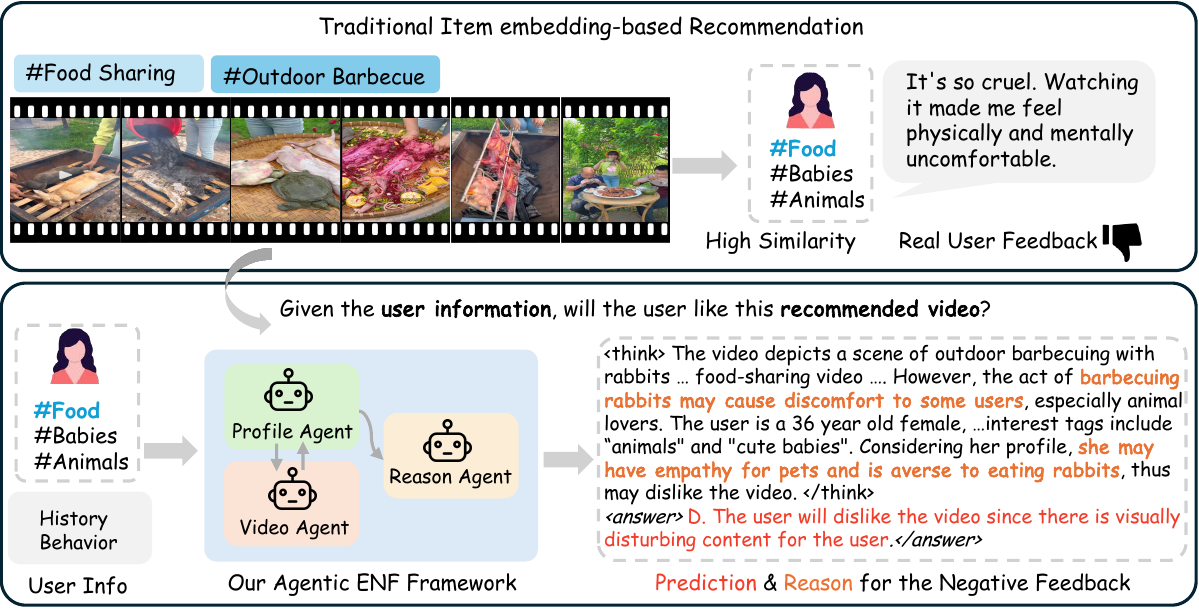}
    
      \caption{\textbf{User negative feedback in a real recommendation scenario.} Traditional method recommends a food-sharing video to a user who loves food-related themes based on high embedding similarity, while this triggers strong negative feedback. Our ENF framework successfully predicts the reason for user's negative feedback, avoiding similar recommendations in the future.
    }
    \label{fig:intro}
    
\end{figure*}

With the explosive growth of online multimodal content, short video platforms such as TikTok and Kuaishou have become primary channels for people to entertain themselves, shop, and access key information, exerting a significant impact on our daily life.
Traditional recommendation methods~\cite{huang2015tencentrec,zheng2018spectral,ying2018graph,yang2020mixed,yuan2020parameter},
like collaborative filtering and multimodal methods,
have been widely adopted for their simplicity and scalability. 
However, existing systems inevitably encounter negative user feedback, which involves both explicit forms (\textit{e.g.}, dislikes) and implicit ones (\textit{e.g.}, fast skips). Addressing such negative feedback is crucial for recommendation systems, as it directly reflects users' dissatisfaction and potential misalignment between recommended content and actual preferences. Therefore, we raise a fundamental question:
\textit{Why do some top-ranked recommendations consistently trigger negative user feedback?}

As a key topic in recommendation systems, research on user negative feedback faces three core challenges.
\textbf{First, the scarcity of negative feedback data limits the availability of high-quality datasets for in-depth analysis}.
While explicit signals (\textit{e.g.}, dislikes, comments) are highly informative, they are extremely sparse (accounting for approximately 0.3\% of all interactions). In contrast, implicit feedback \textit{(e.g.}, watch time, skips) is abundant but low-informative and noisy, making the effective utilization challenging.
\textbf{Second, the specific reasons behind users' negative feedback are highly unaddressed}.
Prior methods typically derive dislike-related features by clustering diverse negative feedback signals, then use these features to suppress similar recommendations~\cite{xie2021dfn,wang2023learning,lai2025dar}, without understanding the specific reasons behind dislike, these methods may lead to poor generalization across different scenarios. 
For example, in Fig. \ref{fig:intro},
if a user dislikes a food-related video, it would be inappropriate to suppress all food-sharing recommendations without understanding the actual cause. 
%
\textbf{Third, existing LLM-based methods for negative feedback lack detailed evaluation in multimodal scenarios}.
Although several LLM-based methods~\cite{bao2023tallrec,MLLM-MSR,zhang2025llm} can predict user preferences, they largely overlook the complex multimodal content of items. Moreover, the absence of evaluation regarding explainable reasons significantly undermines their credibility.

To address these gaps, 
we first construct \textbf{TVNF}, a multimodal video recommendation dataset containing diverse negative feedback from the business scenario (\textit{i.e.}, Tencent News).  
It includes basic user profiles and multi-grained interaction data, such as watch time, dislikes, and actual feedback reasons, enabling comprehensive analysis of user behaviors.
Based on specific user feedback contents, we categorize dislike reasons into following types:
negative events, vulgar or conflicting values, boring plots, and visually disturbing elements.
Further,
we propose an MLLM-based agentic framework designed to simulate personalized user perspectives, aiming to understand videos and generate \textbf{E}xplainable diagnostics for \textbf{N}egative \textbf{F}eedback (\textbf{ENF}).
Specifically, the ENF framework comprises three hierarchically structured agents:
(1) The\textbf{ Profile Agent} constructs dynamically updated user profiles by analyzing demographic data and historical viewing patterns, generating additional psychographic features (\textit{e.g.},``passion for fantastic plots").
(2) The \textbf{Video Agent} leverages the multimodal capabilities of MLLMs to decompose video content, providing content descriptions and value analysis to the Profile Agent for cross-modal validation.
(3) The \textbf{Reason Agent} evaluates videos from the user’s perspective using the updated profile from the Profile Agent, ultimately predicting user preference likelihood and generating interpretable explanations.
In addition, to ensure the ENF pipeline to simulate complex human behaviors, 
we propose \textbf{S-GRPO}, a reinforcement learning paradigm which employs a stepwise reward mechanism to address challenging tasks in a progressive manner.
Specifically, it includes three sequential rewards: a binary judgment reward, a multi-choice selection reward, and an interpretability-oriented reward. 
Notably, the latter reward is only triggered when the output of the previous step is correct,
which stands in contrast to earlier GRPO tasks~\cite{liu2025visual, li2025videochat-r1} that rely on a single reward signal per iteration. 
Through this stepwise reward design, the model can incrementally learn and infer human viewing patterns, thereby achieving more accurate predictions.

Experimental results validate the effectiveness of our framework:
compared with MLLMs and state-of-the-art methods that directly utilize reinforcement learning, 
our approach achieves higher accuracy in both fast-skip prediction and explainable reason classification,
effectively bridging the gap between system rankings and actual user preferences.
Our contributions can be summarized as follows:

\begin{itemize}
\item
To the best of our knowledge, we are the first to identify, explain, and evaluate implicit negative feedback using LLMs. We construct TVNF, a practical multimodal video recommendation benchmark with multi-granularity explicit and implicit feedback. It integrates multimodal content, annotated explainable reasons, and labels, enabling evaluation for explainable negative feedback.

\item 
We propose the Agentic ENF framework, which leverages collaborative MLLM-based agents to effectively simulate user behaviors. In addition, we introduce the S-GRPO training strategy, a progressive reinforcement learning paradigm that ensures explainable and personalized video recommendations. 

\item
Extensive experiments demonstrated the improvements by our ENF method in negative feedback prediction and reason explanation. Additionally, we evaluate ENF in real-world recommendation scenario 
and observe 6.2\% improvement in average play rate and 9.4\% decline for fast-skip rate, significantly enhancing user satisfaction.
\end{itemize}

\section{Related Work}

\paragraph{LLM as User Simulator.}
Considering the powerful semantic understanding and reasoning capabilities of LLMs, numerous studies have leveraged them to facilitate user inference simulations~\cite{ma2022visual,wang2023rethinking,zhang2024usimagent,zhang2024agent4rec,feng2025follow,gubs}.
For instance, iEvaLM~\cite{wang2023rethinking} explores two interaction types within a conversational recommendation benchmark: attribute-based question answering and free-form chit-chat using ChatGPT~\cite{achiam2023gpt}.
To simulate user search behavior, USimAgent~\cite{zhang2024usimagent} prompts an LLM-based agent to construct complete search sessions, including querying, clicking, and stopping behaviors, according to specific search tasks.
Agent4Rec~\cite{zhang2024agent4rec} initializes LLMs as agents with unique user profiles that encompass tastes and social traits to simulate more realistic user behaviors.
Additionally, LLM\_Simulator~\cite{zhang2025llm} simulates user preferences by matching the positive and negative attributes of items with LLM-generated user preferences to determine whether a user would like an item. 
However, prior LLM-based user simulation approaches have relied on frozen LLMs, and using them solely through prompting would risk discrepancies with real user behavior and potential hallucinations~\cite{zhang2024agent4rec}.

\begin{figure*}[t]
    \centering

    \includegraphics[width=0.93\linewidth]{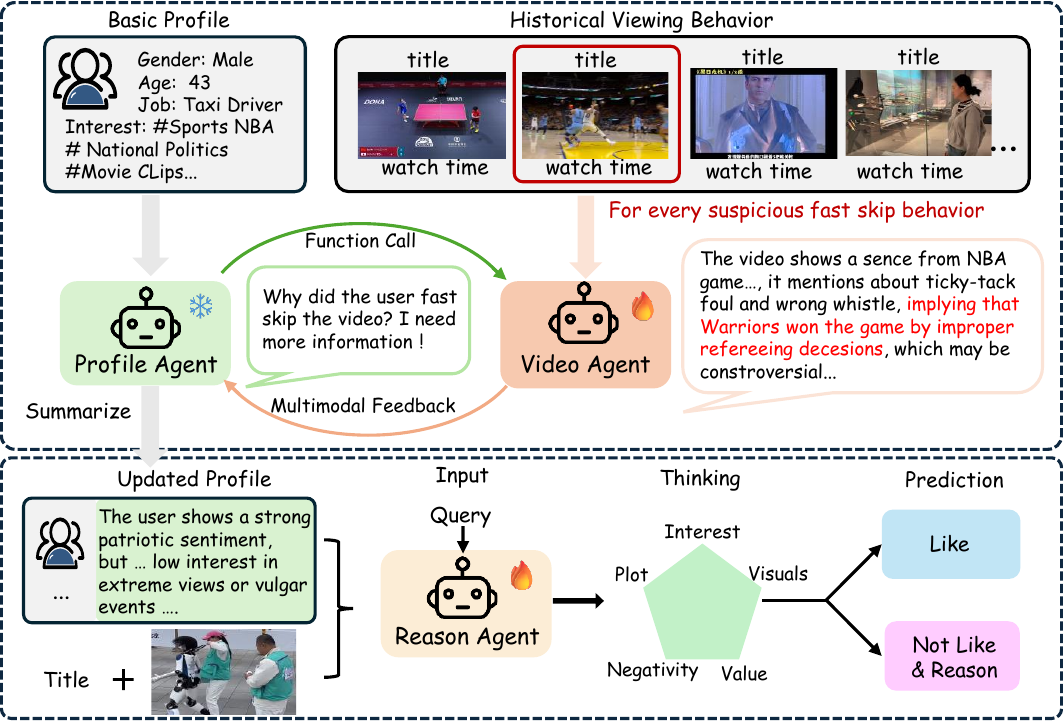}
    \caption{\textbf{Overview of our Agent-based ENF framework.} 
    The three agents collaborate together,
    the Profile Agent analyze the user behaviors to get more comprehensive profile, the Video Agent aids in providing multimodal insights, and the Reason Agent uses the updated profile to predict whether a user likes the recommended video and provides explainable reasons.
    }
    \label{fig:agent}
\end{figure*}

\paragraph{Implicit Negative User Feedback.}
Implicit feedbacks are ubiquitously generated during user browsing behavior, such as video watch rates and impression click-through rates.
They are noisy with subtle signals but still beneficial for recommendation systems~\cite{frolov2016fifty,lai2024adaptive,seo2022siren,cena2023deal,yang2025reccot},
how to effectively leverage such pervasive yet inconspicuous implicit feedback to enhance user understanding persists a critical challenge.
Some works~\cite{pan2016mixed} aim to establish relationships between explicit and implicit feedback through feature mapping or transfer learning. 
DFN~\cite{xie2021dfn} utilizes both internal and external feedback interactions to learn users' unbiased preferences for Click-Through Rate prediction.
CDR~\cite{chen2021curriculum} employs users' explicit dislike signals as a metric to evaluate the relative significance of different behavioral sequences.
SINE~\cite{pan2023understanding} models passive-negative feedback as a mismatch of specific sub-interests.
However, all previous methods 
focus solely on item relevance rather than exploring the specific causes of negative feedback, making it difficult to generalize to new items.

\section{Method}

\subsection{Dataset Construction}
We constructed a dataset, named TVNF, from real-world multi-modal video recommendation scenarios on Tencent News. 
It comprises approximately 10,000 users, 20,000 videos and 320,000 interaction behaviors over seven consecutive days. 
For each user, we gathered basic demographic information (age, gender, occupation), interest tags, and detailed viewing behaviors, including video titles, durations and watch times. 
All personally identifiable information was anonymized for privacy protection. 
\begin{table}[h]
    \centering
    \resizebox{1.0\linewidth}{!}{
    \begin{tabular}{c|ccccc}
    \toprule
    Dataset & \begin{tabular}[c]{@{}c@{}}Multimodal\\Video Data  \end{tabular} & \begin{tabular}[c]{@{}c@{}}Explicit\\Feedback  \end{tabular} & \begin{tabular}[c]{@{}c@{}} Implicit\\Feedback  \end{tabular} &  \begin{tabular}[c]{@{}c@{}} Real User\\Dislike Reason  \end{tabular}  \\
    \midrule
  Amazon   &  \XSolidBrush & \Checkmark & \XSolidBrush &  \Checkmark  \\ %
  Yelp  &  \XSolidBrush & \Checkmark & \XSolidBrush &  \Checkmark \\
  MultiFeed &  \XSolidBrush & \Checkmark & \Checkmark &  \XSolidBrush \\ 
  KuaiRand & \XSolidBrush &  \Checkmark &  \Checkmark  & \XSolidBrush \\ 
  MovieLens  & \XSolidBrush & \Checkmark & \XSolidBrush &  \XSolidBrush \\ 
  MircoLens &  \Checkmark & \XSolidBrush  & \XSolidBrush & \XSolidBrush \\
     \midrule
  TVNF & \Checkmark & \Checkmark & \Checkmark & \Checkmark \\ 
    \bottomrule

    \end{tabular}
    }
   
    \caption{\textbf{Content comparison with previous datasets.}}
    \label{table:data}
    
\end{table}
To ensure data quality, we applied a filtering criterion to ensure that users have at least 15 recorded viewing instances. 
Each of the 20,000 unique videos is accompanied by its original URL and 16 uniformly sampled frames for visual content analysis. 
Additionally, a distinctive advantage of our interpretable benchmark is that
we have collected approximately 1k specific instances of users' negative feedback (such data is extremely scarce and difficult to collect in practice).
In these instances, users explicitly state their reasons for disliking specific videos, which serve as important references for understanding preference mismatches.
By analyzing real user comments, we categorize the reasons for negative feedback into the following types:
i) the video contains negative events, vulgar content, or conflicting values for the user;
ii) the video content lacks sufficient appel and fails to arouse the user's interest; 
iii) the video contains disturbing visual elements that cause discomfort to the user.
For the remaining vast volume of data, we treat cases where the user's viewing rate is below 0.3 as implicit negative feedback.
We instruct GPT-4o~\cite{gpt4o} to label each implicit negative feedback instance according to the aforementioned categories, followed by manual verification to ensure more reliable classification.
Content comparison with previous datasets~\cite{hou2024amazon,asghar2016yelp,xie2021dfn,gao2022kuairand,harper2015movielens,ni2023content}
are presented in Tab.~\ref{table:data}.

\subsection{Our Agent-based ENF Pipeline}
In contrast to objective questions with standard answers, user behaviors are highly subjective and individual, rendering the simulation of specific user behaviors complex. 
Consequently, we propose a multigrained agent-based framework to collaboratively perform prediction and causal analysis of user behaviors, as shown in Fig.~\ref{fig:agent}.

First, the Profile Agent infers users' psychological profiles from their behavioral patterns,
with the objective of addressing key limitation of traditional recommendation systems that rely solely on interest tag embeddings.
Traditional systems, for instance, often neglect nuanced psychological tendencies of users:
a celebrity fan may strongly prefer positive content about their idol while rejecting critical narratives; a food enthusiast might react negatively to videos with overly graphic depictions of ingredient preparation.
These examples highlight that user preferences extend beyond surface-level interest tags, being rooted instead in deeper psychological traits.
Thus, the core goal of the Profile Agent is to uncover these latent psychological traits, enabling more nuanced and user-aligned recommendations.
To achieve this, the agent leverages basic user profile information (age, gender, occupation, and recent interests) to analyze users' sequential watching behaviors, including titles and play rates.
Here, the agent focuses on videos with a play rate below 0.3,
which indicate user dissatisfaction with recommendations. 
When textual titles alone provide insufficient clues, the Profile Agent dynamically activates the Video Agent to extract multimodal clues, enhancing the depth of this analysis.

The Video Agent then conducts in-depth analysis at the individual video level.
Beyond basic content description, it identifies potentially controversial elements within the video and provides contextual explanations. 
Concurrently, through detailed analysis of each interaction, the Profile Agent iteratively identifies factors causing users to skip content and updates psychological profiles (\textit{e.g.}, value orientations and tolerance for negativity toward such videos).
Finally, the Reason Agent leverages basic user information and refined psychological profiles to generate user-centric video understanding. 
This analysis process encompasses four key dimensions:
whether the video's content aligns with the user's interest;
whether the plot is appealing;
whether the content contains negative events or extreme opinions;
and whether the visual elements align with the user's sensory tolerance.
By assessing these factors, the Reason Agent infers the user's attitude toward the video.

\begin{figure*}
    \centering
   
    \includegraphics[width=0.95\linewidth]{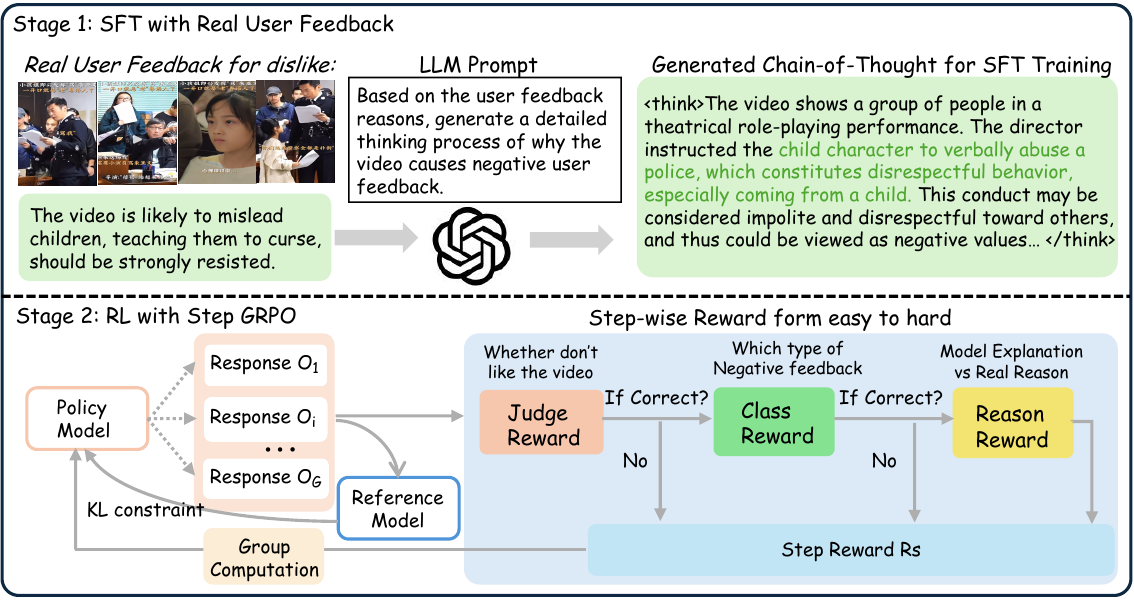}
  \caption{\textbf{Training process of our agents.} In the first stage, we use real user feedback reasons for cold start; and in the second stage, we propose a progressive reward mechanism that provides step rewards for a response in an order from easy to hard.}
    \label{fig:step}
    
\end{figure*}

\subsection{Progressive Training Strategy for Agents}
We adopt Qwen2.5-VL-7B~\cite{qwen2math} as the foundational MLLMs 
and follow the two-stage training framework of DeepSeek-R1~\cite{guo2025deepseek}: supervised fine-tuning (SFT) for cold start, followed by reinforcement finetuning (RFT). 
Specifically,
in Stage 1, leveraging reasons of real user feedback, we prompt GPT-4o to generate a chain-of-thought (CoT) reasoning process explaining why users disliked specific videos. These generated CoTs of the reasons are used as SFT data to warm up the model.
In Stage 2, we further train the agent on the non-annotated data using our proposed S-GRPO algorithm.
Our task definition involves a hierarchical prediction framework consisting of three progressive stages:
first, binary judgment to determine whether the user generates negative feedback; 
second, multi-choice classification of pre-defined negative feedback types; 
and third, generating reasons to explain the core cause of the negative feedback.
Unlike prior methods~\cite{video-r1,li2025videochat-r1,wang2025videorft} that only involve a single objective question for a video, 
our method introduces multiple granularity judgments, which makes it challenging to evaluate the response using a single reward.

To address this issue, we propose an effective multiple-choice question paired with Step Group Relative Policy Optimization (S-GRPO). 
Except the basic format reward,
this design incorporates a progressive reward mechanism that offers three step rewards ${R_S}_i$ for a response $o_i$,
as depicted in Figure~\ref{fig:step}.
In the binary Judge Reward $r_{\text{judge}}$ at the first step, we determine whether the prediction of the user's attitude is correct. 
If the judgment is wrong, the process terminates immediately. 
Otherwise the response will receive a fixed reward (\textit{e.g.}, 0.5),
and if the real user feedback is positive, the process also terminates,
and if the feedback is negative,
the process proceeds to the second step, i.e., the Class Reward $r_{\text{class}}$. 
Here, if the choice of the negative feedback type is accurate, an additional reward (\textit{e.g.}, 1.0) is granted, and the process advances to the third step: the Reason Reward $r_{\text{reason}}$. 
At this stage, we calculate the average of the ROUGE-1, ROUGE-2, and ROUGE-L scores between the reasoning content within the \textit{$<$think$>$} tag and the actual user feedback reasons, which is then used as an extra reward.
Note that the Video Agent is trained using 3-step rewards on explicit negative feedback data,
while the Reason Agent is trained with 2-step rewards on implicit data due to the lack of ground truth reasons.
This progressive design encourages the model to tackle problems from easy to hard.
For instance,
it allows the model to earn rewards even when the multiple-choice answer is wrong but the binary judgment is correct;
meanwhile, correct classifications accompanied by sound reasoning processes are assigned higher scores.
Additionally, the advantage of $A_i$ of response $o_i$ among $G$ responses is computed based on the rewards within each group.
Building on this framework, the model gradually learns to classify and explain the underlying reasons, with the final policy updated to maximize the objective as follows:
\begin{equation}
\resizebox{0.38\linewidth}{!}{%
$\begin{aligned}
    A_{i} = \frac{R_i - \mathrm{mean}(\{R_j\})}{\mathrm{std}(\{R_j\})}
    \end{aligned}$%
}
\end{equation}
\begin{equation}
\resizebox{\linewidth}{!}{%
$\begin{aligned}
    \mathcal{J}_{GRPO}(\theta) =& \min \left( \frac{\pi_\theta(o_i | q)}{\pi_{\theta_{old}}(o_i | q)} A_i, \mathrm{clip} \left( \frac{\pi_\theta(o_i | q)}{\pi_{\theta_{old}}(o_i | q)}, 1 - \varepsilon, 1 + \varepsilon \right) A_i \right) \\
    & - \beta \mathcal{D}_{KL}(\pi_\theta || \pi_{ref}),
\end{aligned}$%
}
\end{equation}

\section{Experiments}
\begin{table*}[t]
\centering

\small
\begin{tabular}{ll|c|cccc|cc}
\toprule[1pt]
& \multirow{2}{*}{\textbf{Model}} 
 & \multirow{2}{*}{\textbf{Size}} & \multicolumn{4}{c|}{\textbf{Binary Judgment}} & \multicolumn{2}{c}{\textbf{Explain}} \\
& & & \textbf{Acc} & \textbf{Precision} & \textbf{Recall} & \textbf{F1\_Score} & \textbf{Class\_Acc} & \textbf{Reasoning} \\
\midrule

& GPT-4o~\cite{gpt4o} & -& \textbf{0.882}   &  \underline{0.895}   &  \underline{0.630}  &  \underline{0.739} &  \underline{0.568} & \underline{0.402}  \\
& DeepSeek~\cite{guo2025deepseek} &- & 0.849 & \textbf{0.917} &  0.440 &  0.594  &  0.352 & 0.266  \\
& LLaMa~\cite{grattafiori2024llama} & 70b  & 0.822 & 0.900 & 0.346 &   0.500 &  0.268   &  0.177      \\
& Qwen2.5VL~\cite{yang2024qwen2} &   7b    & 0.815 & 0.816 & 0.423 & 0.564 & 0.296 & 0.229 \\
& Qwen3~\cite{yang2025qwen3} & 32b & 0.845 & 0.833 & 0.464 & 0.596 & 0.346 &  0.238    \\

\midrule
&Video-R1~\cite{video-r1} & 7b & 0.835 & 0.789 & 0.540 & 0.667 & 0.432  & 0.318 \\
& VideoChat-R1~\cite{li2025videochat-r1} & 7b &  0.842 & 0.850 & 0.654 & 0.739 &  0.500 & 0.383                              \\
\rowcolor{gray!20}
& Our Video Agent & 7b & \underline{0.861} & 0.706 & \textbf{0.808} & \textbf{0.750} & \textbf{0.654} & \textbf{0.537}    \\

\bottomrule[1pt]
\end{tabular}

\caption{\textbf{Prediction of explicit user negative feedback: understanding and reasoning about video controversy content.}
Recall serves as a key metric for the recognition rate of negative feedback videos.
The reasoning score is calculated using real user feedback reasons only when the judgment is correct.
}
\label{table:video}
\end{table*}

\begin{table*}[t]
\centering
\small
\begin{tabular}{ll|c|cccc|c}
\toprule[1pt]

& \multirow{2}{*}{\textbf{Model}} 
 & \multirow{2}{*}{\textbf{Size}} & \multicolumn{4}{c|}{\textbf{Binary Judgment}} & \multicolumn{1}{c}{\textbf{Explain}} \\
& & & \textbf{Acc} & \textbf{Precision} & \textbf{Recall} & \textbf{F1\_Score} & \textbf{Class\_Acc} \\
\midrule

& GPT-4o~\cite{gpt4o} & - & 0.575   &  \underline{0.396}   &  \textbf{0.796}  &  \underline{0.521} &  0.502  \\
& DeepSeek~\cite{guo2025deepseek} & - &  \underline{0.608} & 0.331 & 0.626 & 0.433   &  0.476   \\
& LLaMa~\cite{grattafiori2024llama} & 70b & 0.601 & 0.303 & 0.233 & 0.264 & 0.155   \\
& Qwen2.5VL~\cite{yang2024qwen2} & 7b & 0.528 & 0.359 & 0.733 & 0.482 & 0.435       \\
& Qwen3~\cite{yang2025qwen3} & 32b & 0.548   &  0.368   &  0.708   & 0.484  & 0.425     \\

\midrule
& SASRec~\cite{kang2018self} & - & 0.448  &   0.230    &  0.358    &  0.279    &    -  \\
& MLLM-MSR~\cite{MLLM-MSR} & 7b & 0.545 & 0.355 & 0.632 & 0.455 & -   \\
& Video-R1~\cite{video-r1} & 7b  & 0.573 & 0.390 & 0.748   & 0.513 &  0.494   \\
& VideoChat-R1~\cite{li2025videochat-r1} & 7b & 0.561 & 0.384 & 0.775 & 0.516 & \underline{0.512}  \\    
\rowcolor{gray!20}
& Our ENF & 7b &  \textbf{0.612} & \textbf{0.404} & \underline{0.782} & \textbf{0.533} & \textbf{0.543}   \\

\bottomrule[1pt]
\end{tabular}
\caption{\textbf{Prediction of implicit user negative feedback: simulating user video-watching behavior.} Despite the greater difficulty of implicit behavior prediction due to significant noise, our method still achieves the best results.} 
\label{table:user}
\end{table*}

\subsection{Implementation Details and Metrics}
We conducted experimental validation on two granularities of negative feedback using the collected dataset: explicit negative feedback with real users' dislike reasons, and implicit negative feedback derived from users' fast-skip data. 
For the explicit data, negative feedback videos are mixed with normal videos, and the Video Agent is used to predict whether a video contains controversial content that may trigger negative feedback, as well as providing explanations.
For the implicit data, following the sequence recommendation setting, the Reason Agent predicts whether a specific user would generate negative feedback for a recommended video based on their historical behaviors.
In our implementation, GPT-4o serves as the Profile Agent, while Qwen-2.5VL-7b works as both the Video Agent and Reason Agent.
We adopt full-parameter fine-tuning:
the Video Agent and Reason Agent are trained with 4 80G GPUs using the constructed dataset, 
which contains approximately 2,000 instances for training and evaluation.
We use 16 images and the video title as the video feature inputs,
the group size $G$ is set to 8 and the learning rate is 1e-6.
Evaluation metrics include binary accuracy, precision, recall, and F1-score for negative feedback prediction, as well as the accuracy of feedback reason classification. 
For explicit negative feedback, we employ GPT-4o to assess the relevance between the model's explanations and users' real feedback reasons, with relevance scores ranging from 0 to 1.

\subsection{Main Results}

We next present predictions for both users' explicit and implicit negative feedback. 
Explicit feedback is typically triggered by overtly controversial contents in videos, which could not be recognized by traditional embedding-based methods. Therefore, we employ the Video Agent to analyze concrete multimodal content directly.
These negative videos are mixed with randomly selected normal videos at an approximate ratio of 1:4, and the Video Agent is tasked with predicting whether a video is potentially controversial, along with providing explanations.
The results are shown in Tab. \ref{table:video}. In terms of video content understanding, while GPT-4o achieves the highest prediction accuracy, existing models generally lack sensitivity to controversial content, leading to low recall rates for negative feedback videos. In contrast, our method, leveraging the S-GRPO approach that learns from easy to difficult tasks, attains the highest Recall (0.808) and F1 Score (0.750). Although the increased recall slightly reduces precision, the method also delivers the most accurate results in both reason classification and explanation (with +8.6\% and +13.5\% improvements over GPT-4o, respectively). This capability is critical for identifying problematic videos and mitigating poor user experiences.

In terms of implicit negative feedback prediction, we adopt a personalized user perspective. By analyzing user profiles and historical behaviors, we infer their psychological preferences and employ the Reason Agent to predict users' attitudes toward videos (\textit{e.g.}, whether they will fast-skip). 
As shown in Tab. \ref{table:user}, predicting implicit feedback is far more challenging than explicit feedback: the highest accuracy is only 61.2\%, and precision rates are generally low. This may stem from real-world user behaviors being influenced by multiple factors with significant noises, resulting in inherent randomness.
We also evaluated traditional methods such as SASRec~\cite{kang2018self},
however, these methods exhibit poor discriminative performance in cold-start scenarios that require fine-grained item differentiation.
In addition,
among LLM-based methods, 
GPT-4o achieves the highest recall rate, but directly applying such large models in a zero-shot setting will yield unsatisfactory performance.
For the training of smaller 7b models, our ENF framework outperforms previous video-reasoning methods in both fast-skip prediction and reason classification, validating its effectiveness.
This ability to attribute reasons for implicit negative feedback significantly enhances user intent understanding, not only aiding in identifying issues and improving existing recommendation systems, but also advancing next-generation explainable recommendations.


\subsection{Ablations}

In this section,we conduct an ablation study by designing different variants of our models.
For the Video Agent, we ablate the training of our generated CoT process as SFT data, 
the RL process and our proposed S-GRPO reward.
As shown in Tab.~\ref{table:abvideo}, all ablated models perform worse than the full Video Agent across all benchmarks.
Specifically, the SFT process enables the model to acquire user-side prior knowledge for cold-start, removing this component leads to a noticeable performance drop, particularly in prediction accuracy.
The RL process further encourages the model to think deeply and diversely, thereby enhancing overall performance.
Moreover, without our progressive S-GRPO training, the model tends to focus solely on binary judgment and struggles to distinguish between classification tasks and the underlying reasons for controversial factors.

For the Reason Agent, we ablate three designs: the Profile Agent, the initialization with the Video Agent, and the S-GRPO mechanism. As presented in Tab.~\ref{table:abuser}, the Profile Agent provides richer psychographic features for more comprehensive user modeling; the Video Agent offers empirical priors learned from explicit user negative feedback, which aids in predicting implicit negative feedback; and S-GRPO ensures performance balance between binary judgment and classification-reasoning tasks. All modules collectively contribute to the improvement of the final performance.

\begin{table}
    \centering
    \small
    \setlength{\tabcolsep}{1mm}
    \begin{tabular}{ccc|cccc}
    \toprule
    SFT & RL & S-GRPO &  Acc   & F1\_Score  & Class\_Acc  & Reasoning\\
    \midrule
     \XSolidBrush & \XSolidBrush & \XSolidBrush &  0.815  & 0.423   & 0.296  &  0.229 \\ 
      \XSolidBrush & \Checkmark & \Checkmark & 0.830   & 0.686   &  0.592     & 0.492  \\ 
    \Checkmark & \XSolidBrush & \XSolidBrush & 0.851   & 0.615    &  0.346    & 0.312  \\ 
     \Checkmark & \Checkmark & \XSolidBrush &  0.845  & 0.667    &   0.412   & 0.339 \\ 
    
    \Checkmark & \Checkmark & \Checkmark &  \textbf{0.861} & \textbf{0.750} & \textbf{0.654} & \textbf{0.537} \\ 
    \bottomrule

    \end{tabular}

    \caption{\textbf{Ablation}. Training Processes on Video Agent.}
    
    \label{table:abvideo}
    
\end{table}

\begin{table}
    \centering
    \small
    \setlength{\tabcolsep}{1mm}
    \begin{tabular}{ccc|ccc}
    \toprule
    \begin{tabular}[c]{@{}c@{}} Profile\\Agent  \end{tabular}  & \begin{tabular}[c]{@{}c@{}} Video\\Agent  \end{tabular} & S-GRPO & Acc   & F1\_Score  & Class\_Acc \\
    \midrule
    \XSolidBrush & \XSolidBrush & \XSolidBrush & 0.528 & 0.482 & 0.435 \\
    \XSolidBrush & \Checkmark &\Checkmark & 0.596  &  0.518   & 0.508  \\
     \Checkmark & \XSolidBrush & \Checkmark &  0.573  &   0.513  &  0.504     \\ 
     \Checkmark & \Checkmark & \XSolidBrush & 0.535   & 0.488    &  0.522      \\ 
    \Checkmark & \Checkmark & \Checkmark &  \textbf{0.612}  &  \textbf{0.533}   & \textbf{0.543}   \\
    \bottomrule

    \end{tabular}
   
    \caption{\textbf{Ablation}. Training Processes on Reason Agent.}
    \label{table:abuser}
\end{table}

\begin{table}[t]
\centering
\small
\begin{tabular}{c|cc|cc}
\toprule
 & \multicolumn{2}{c|}{\textbf{MovieLens}} & \multicolumn{2}{c}{\textbf{Steam}} \\ 
\textbf{Method} & \textbf{Acc} & \textbf{F1\_Score}  & \textbf{Acc} & \textbf{F1\_Score} \\ 
\midrule
GPT-4o & 0.584 & 0.600 & 0.634  &  0.662  \\

RecAgent & 0.581 & 0.621 & 0.627 & 0.650 \\

Agent4Rec & 0.691 & 0.698 & 0.689 & 0.679 \\

SimUSER & 0.791 & 0.777 & 0.791  & 0.794 \\
\rowcolor{gray!20}
Ours & \textbf{0.815}  & \textbf{0.808}  & \textbf{0.803}   & \textbf{0.805} \\
\bottomrule

\end{tabular}

\caption{User preference alignment across MovieLens and Steam datasets.}
\label{table:other}
\end{table}

\begin{table}[h]
\centering
\small
\resizebox{\linewidth}{!}{
\begin{tabular}{l|ccc}
\toprule[1pt]
\multicolumn{1}{c|}{\textbf{Method}} & \textbf{Avg\_Time}$\uparrow$ & \textbf{Fast-skip Rate}$\downarrow$ & \textbf{Dislike Rate}$\downarrow$\\
\midrule
Base RS  & 47.6\% &  23.7\%  & 0.61\% \\
\midrule
\rowcolor{gray!20}
Base RS + ENF & 53.8\%   &  14.3\%  & 0.35\% \\

\bottomrule[1pt]
\end{tabular}
}
\caption{Performance evaluation in the real-world scenario.
}
\label{table:baseline}

\end{table}

\paragraph{Generalization on other Datasets.}
To evaluate the performance of our method in other domains, we further conduct evaluation on MovieLens~\cite{harper2015movielens} and Steam~\cite{kang2018self} following Agent4Rec~\cite{zhang2024agent4rec} to simulate users’ preferences toward items.
Tab.~\ref{table:other} shows that, 
previous methods~\cite{bougie2025simuser,wang2025user} only prompt frozen LLMs for the prediction, which heavily rely on LLM's pre-training performance,
and also suffer from inherent hallucinations.
In contrast, our method achieves higher prediction accuracy through RL-based alignment with user preferences.

\paragraph{Testing on Business Platform.}
We further evaluate ENF in real-world video recommendation scenarios on Tencent News.
For selected users, we use their behavioral data from previous days as the reference and assess performance over the subsequent few days.
The original recommendation system (labeled Base RS) generates candidate videos, 
while our ENF framework predicts user attitudes towards these videos and filters out those likely to trigger negative feedback.
Evaluation metrics include the average watch time, fast-skip rate, and dislike rate of recommended videos. 
As observed, our method yields significant improvements across these three metrics, with 13.0\%, 39.7\% and 42.6\% improvements over the baseline, respectively. This validates the effectiveness of our approach in enhancing user satisfaction.

\section{Conclusion}
In this paper, we conduct research on users' negative feedback in recommendation systems. 
We first construct a benchmark dataset with real reasons for negative feedback,
and propose the ENF Agentic framework with three hierarchically structured agents
for multimodal behavior analysis and interpretable predictions. 
Leveraging the S-GRPO training paradigm, we improve prediction accuracy and achieve reliable explanations to finally improve recommendations.
As a pioneering effort, this work offers novel insights into negative feedback and advances next-generation explainable recommendations, and we hope to inspire more exploration.

\section*{Acknowledgements}
This work was supported by the National Key R\&D Program of China(NO.2022ZD0160505).

\bibliography{aaai2026}
\newpage

\appendix
\newpage

\section{Appendix of the Paper}
This is the appendix of the paper: When Top-ranked Recommendations Fail: Modeling Multi-Granular Negative Feedback for Explainable and Robust Video Recommendation, and we give more details for the discussion of the main paper.

\subsection{A.1 Prompt for the Agents}
In this section, we give the concrete prompts for the instruction of our agents.

For the Profile Agent:
$P_{profile} = $ \textit{'You are a helpful assistant for user behavior analysis. Given the basic user information \textbf{\{age, gender, occupation, and recent interests\}} and historical behavior \textbf{\{watch history\}}, among the watch history, "play\_rate" indicates the user’s video watch completion rate, a low play\_rate means the user may dislike the video. Please focus particularly on videos with low play\_rate. You may request to call the Video Agent to obtain more detailed information about these videos when necessary. Finally, summarize the user’s personality and psychological tags, such as their sensitivity to negative or vulgar content.'}.

For the Video Agent:
$P_{video} = $ \textit{'You are a helpful assistant for video content analysis. The assistant first thinks about the reasoning process in the mind and then provides the answer. Based on the video \textbf{\{title\}} and visual contents, does the video contain any potential controversial element that may cause user uncomfortable? If yes, what type of controversy does it contain? Candidate answers: \textbf{\{candidates\}}. 
The reasoning process and answer are enclosed within $<$think$>$ $<$/think$>$ and $<$answer$>$ $<$/answer$>$ tags, respectively, i.e., $<$think$>$ reasoning process here $<$/think$>$$<$answer$>$ answer here $<$/answer$>$.  
Please provide only the single option letter (e.g., A, B, C, D, etc.) within the $<$answer$>$ tags.'}.

For the Reason Agent:
$P_{reason} = $ \textit{'You are a helpful assistant for video recommendation. The assistant first thinks about the reasoning process in the mind and then provides the answer. Based on the user information \textbf{\{updated profile\}}, will the user like the recommended video \textbf{\{video\}}? Think from the user's perspective to predict whether the user will fast skip the video, if yes, what kind of reason causes the negative feedback? Consider the value, plot, negativity, visual disturbing elements of the video. Candidate answers: \textbf{\{candidates\}}.
The reasoning process and answer are enclosed within $<$think$>$ $<$/think$>$ and $<$answer$>$ $<$/answer$>$ tags, respectively, i.e., $<$think$>$ reasoning process here $<$/think$>$$<$answer$>$ answer here $<$/answer$>$.  
Please provide only the single option letter (e.g., A, B, C, D, etc.) within the $<$answer$>$ tags.'}.

\subsection{A.2 Additional Details for Dataset Construction}
In this section, we give more details of the TVNF dataset.
The user data is collected from real-world scenario logs of the Tencent News APP, with 7 days of data extracted. 
Since it contains a lot of noisy signals, we filtered out some obvious anomalies, such as a play rate greater than 5 (which may be due to prolonged inactivity) and a watching duration of less than 0.5 seconds (which may be due to accidental taps by users).
We follow the metrics to collect the data where:
(1) users performed operations both before and after the feedback; (2) the interval between these actions was within 2 minutes; and (3) the user did not leave the News APP. This ensures the retained data reflects genuine user disliking of the recommended videos. Additionally, videos with fewer than 10 views were excluded (given potential poor quality or short retention time on the platform).
Overall, there are 20,539 videos form various domains, such as sports, health care, entertainment, politics, TV shows, etc..
The duration of the videos ranges from 5 seconds to over 10 minutes, videos with a duration of 10–30 seconds account for the highest proportion, with an average duration of about 46 seconds.
And these selected videos are ensured to have been viewed at least 10 times.

\subsection{A.3 Additional Training Details}
In this section, we give more details for the training process.

\begin{figure}[h]
    \centering

    \includegraphics[width=\linewidth]{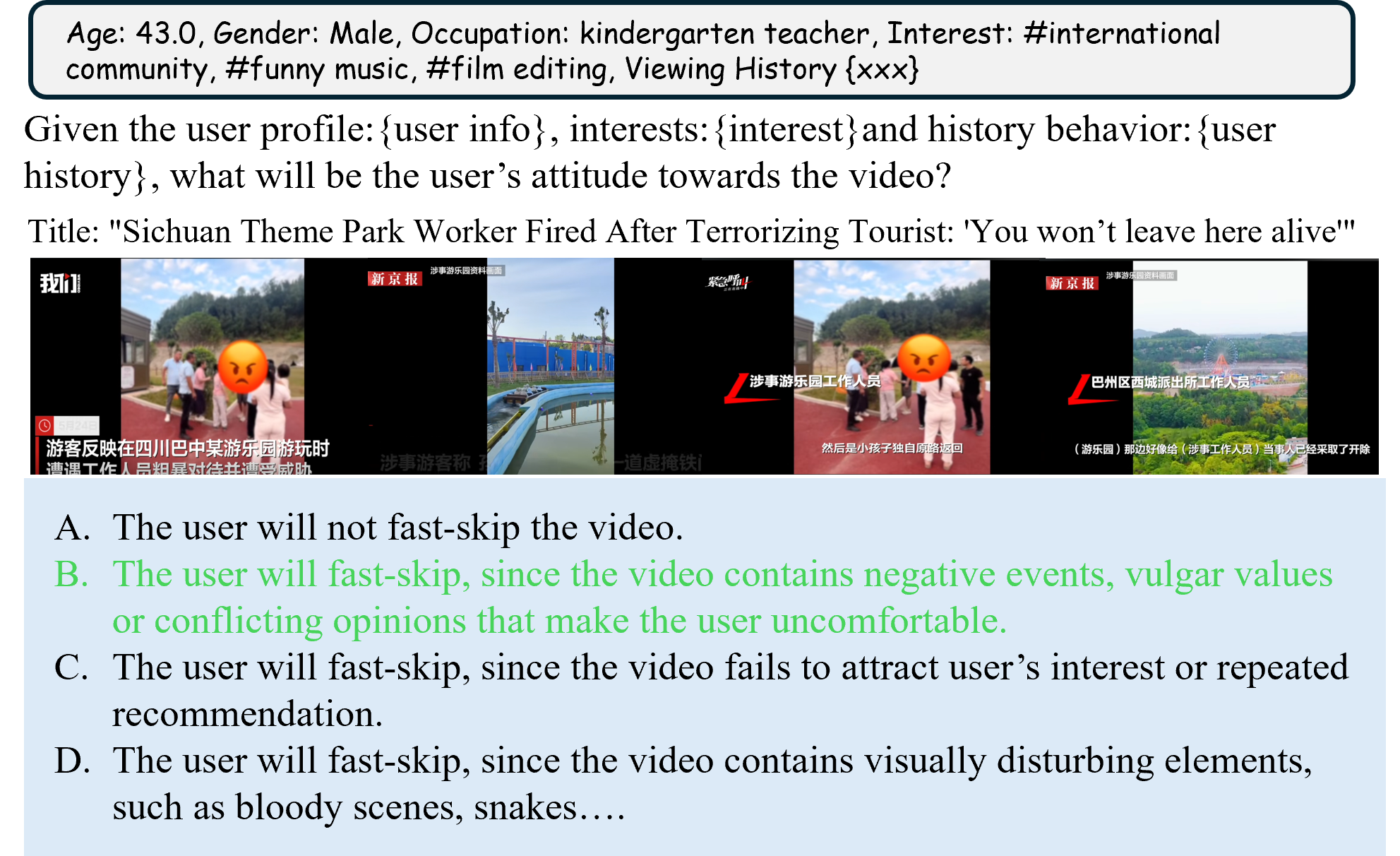}
    \caption{Training Samples. The green line means the ground truth answer. }
    \label{fig:example}
\end{figure}

\paragraph{Training / Text dataset details:} For the sequential recommendation task, we select 1,000 user-item instances as the test set, with the remaining user behaviors serving as the training set. 
For each user, the first half of their viewing history is used as reference input context, 
while the latter half of the behaviors are regarded as the ground truth of 
whether the user would fast-skip the video, a concrete example could be seen in Fig.~\ref{fig:example}.
The training pipeline is built on the open-rl framework with an efficient codebase, running on 4 NVIDIA A100 80G GPUs. 
It adopts bf16 precision and integrates Zero2, Flash-Attention, and with VLLM employed for acceleration; 
the maximum generation length is set to 1024. 
We first use the generated SFT data for cold-start for one epoch,
and then conduct RL fine-tuning for three epochs, we use Adam optimizer with a learning rate of 1e-5, a weight decay of 0.01, and a beta value of 0.04.

\begin{figure*}
    \centering

    \includegraphics[width=\linewidth]{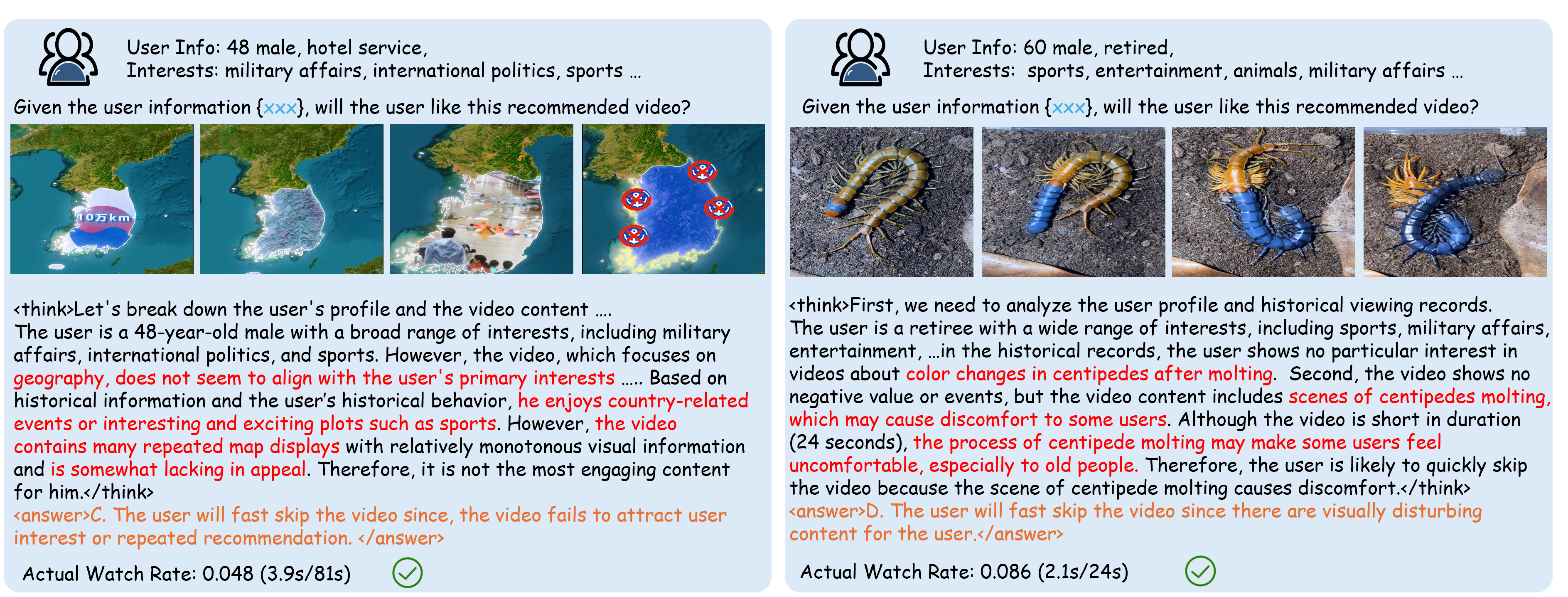}
    \caption{\textbf{Case Study.} We present specific user examples to illustrate why they choose to fast-skip the video.}
    \label{fig:case}
    \vspace{-0.1cm}
\end{figure*}

\paragraph{Basic RL Standards for LLM.}
Without loss of generality, 
we adhere to the standard notations presented in the classic works of reinforcement learning~\cite{sutton1998reinforcement,agarwal2019reinforcement}.
More specifically, we use  
\(s \in \mathcal{S}\) to denote the state space, 
\(a \in \mathcal{A}\) to denote the action space,
\(r_k\) to denote the reward function in step k,
\(\mathcal{P}\) to denote the transition dynamics,
\(\pi(a|s)\) is the probability of performing action \(a\) in state \(s\) under policy \(\pi\),
and \(\gamma \in [0,1]\) is the discount factor.
The goal is to maximize the discounted cumulative returns for each trajectory as below,
\begin{equation}
G_t = \sum_{k=t+1}^{T} \gamma^{k-t}r_k
\end{equation}
where T is the maximum step numbers per episode.
Instead of using the classic PPO~\cite{schulman2017ppo} algorithm that requires a critic model to evaluate policy performance,
we use the GRPO~\cite{shao2024deepseekmath} to compare groups of candidate responses directly.
\begin{equation}
\begin{split}
\mathcal{J}_{\text{GRPO}}(\theta)
&= \mathbb{E}_{[q\sim P(Q), \{o_i\}_{i = 1}^{G}\sim \pi_{\theta_{\text{old}}}(O\mid q)]} 
  \frac{1}{G}\sum_{i = 1}^{G}\frac{1}{|o_i|}\sum_{t = 1}^{|o_i|} \Bigg\{ \\
&\quad \min\bigg[
\frac{\pi_{\theta}^{i,t}}{\pi_{\theta_{\text{old}}}^{i,t}}\hat{A}_{i,t}, 
\text{clip}\left(\frac{\pi_{\theta}^{i,t}}{\pi_{\theta_{\text{old}}}^{i,t}}, 1 - \epsilon, 1 + \epsilon\right)\hat{A}_{i,t}
\bigg] \\
&\quad - \beta\mathbb{D}_{\text{KL}}[\pi_{\theta}\|\pi_{\text{ref}}]
\Bigg\}
\end{split}
\end{equation}
\begin{equation}
    \hat{A}_{i,t}=\frac{r_i - \text{mean}(\textbf{r})}{\text{std}(\textbf{r})}
\end{equation}
Given a problem \(q\) for the model \(\pi_{\theta}\), it samples to generate a group of distinct answers \( {o_i} \), where \(i = 1, 2, \dots, G\), \(G\) is the sampled number in the group. 
Each answer has a different length \(|o_i|\).
\({ \pi_{\theta}^{i,t}} \) is the policy probability of decoding the \(t\)-th token of the sampled answer.
The KL term constrains that the distribution of \(\pi_{\theta}\) should not deviate too much from the original policy \(\pi_{\text{ref}}\) by penalty coefficient \(\beta \). 
Here, an optimized KL term is adopted, which has the characteristics of being unbiased and having a small variance.
The clip strategy restricts the ratio between $\frac{\pi_{\theta}}{\pi_{\theta_{old}}}$, and by limiting the ratio within the interval $\varepsilon$, it prevents the new strategy from having large numerical updates. 
\(\textbf{r} = \{r_1, r_2, \dots, r_G\}\), and \(\hat{A}_{i,t}\) is the relative advantage of the \textit{i}-th answer.
Through the optimization of \(\mathcal{J}_{\text{GRPO}}(\theta)\),
GRPO encourages the model to choose the answer with higher reward within the group.

\paragraph{Deployment Details and Efficiency}
To improve response speed, we deployed our ENF with FP16 quantization. Specifically, ENF can analyze 1000 users within 15 minutes through the asynchronous invocation mechanism, which could help to process about one query per second.

\subsection{A.4 Case Study}
In this subsection, we present case studies to further demonstrate the effectiveness of our negative feedback reasoning framework.
Two examples of real user video-watching behaviors are showcased in Fig.~\ref{fig:case}.
Through in-depth analysis of video content and personalized user behaviors, our model successfully predicts user dislikes, even for unseen video items in cold-start scenarios. Furthermore, by learning priors learned from explicit user negative feedback reasons, the model provides reasonable explanations for fast-skip behaviors.
For example, it accurately identifies that a 48-year-old male, who prefers engaging or thrilling plot content, may find geographical knowledge content boring and unappealing.
And the centipedes molting scene may be visually disturbing to a man who likes entertainment.

\subsection{A.5 Evaluation on Other Datasets}
In this section,  we verify the effectiveness of our method across different domains, i.e., moves dataset MovieLens-1M~\cite{harper2015movielens} and game dataset Steam~\cite{kang2018self}.
Following previous setting of user simulation agents Agent4Rec~\cite{zhang2024agent4rec} and SimUser~\cite{bougie2025simuser},
we query the agents to classify whether the user would like a specific item,
based on their their actual behaviors, the items that user has interacted with high ratings are regarded as positive, otherwise negative.
And we select 1,000 instance agents each with 20 candidate items (positive and negative in 1:1 setting), and perform the evaluation as a binary classification task.
Note that, these datasets do not involve multimodal information, so we don't consider the video agent, only use textual descriptions and train the agent with RL.
Previous methods~\cite{zhang2024agent4rec,sen,wang2025user,lvagent,vragent,videochata1,pcaby,yue2025uniflow} only prompt frozen LLMs for the prediction, it heavily relies on the LLM's pre-training performance,
and also suffers from inherent hallucinations.
On the contrary, our method achieves higher prediction accuracy through RL-based alignment with user preferences.


\end{document}